\documentclass[12pt]{article}
\usepackage[T2A]{fontenc}
\usepackage[koi8-u]{inputenc}
\usepackage[english]{babel}
\usepackage[dvips]{graphicx}
\usepackage{mathtext}
\usepackage{amsmath}
\usepackage{amssymb}

\begin{document}

\title{Large-scale structure formation in cosmology with classical and tachyonic scalar fields}
\author{O.~Sergijenko$^{(1)}$, Yu.~Kulinich$^{(1)}$, B.~Novosyadlyj$^{(1)}$, V.~Pelykh$^{(2)}$}
\maketitle

\medskip

\centerline{$^{(1)}$ Astronomical Observatory of Ivan Franko National University of Lviv}
\centerline{$^{(2)}$ Pidstryhach Institute of Applied Problems in Mechanics and Mathematics of NASU}

\vskip0.5cm

{\it The evolution of scalar perturbations is studied for 2-component (non-relativistic matter and dark energy) cosmological models at the linear and non-linear stages. The dark energy is assumed to be the scalar field with either classical or tachyonic Lagrangian and constant equation-of-state parameter $w$. The fields and potentials were reconstructed for the set of cosmological parameters derived from observations. The comparison of the calculated within these models and observational large-scale structure characteristics is made. It is shown that for $w=const$ such analysis can't remove the existing degeneracy of the dark energy models.}

\vskip0.5cm
\section*{Introduction}
The observations of the last decade surely confirm the acceleration of the cosmological expansion. The explanation of this fact needs the assumption that the main part -- approximately $70\%$ -- of the energy density of the Universe belongs to the mysterious repulsive component called ``dark energy''. The simplest model describing satisfactory almost the whole set of the experimental data is $\Lambda$CDM-one. Here dark energy is identified with the $\Lambda$-term in the Einstein equations. However, in this case there are several interpretational problems, which suggest that another solution should be found. The most popular alternative approaches are quintessential scalar fields, i.e. scalar fields with the equation-of-state (EoS) parameter $-1<w_{de}\equiv p_{de}/\rho_{de}<-1/3$. The simplest physically-motivated Lagrangians are the classical and tachyonic ones. The first of them is the simple generalization of the non-relativistic particle Lagrangian to the field while the second (called also the Dirac-Born-Infeld one) -- of the relativistic particle one \cite{bagla2003,gibbons2003,padmanabhan2002,roy-choudhury2002,sen2003}. The Lagrangian of classical field has the canonical
kinetic term, the Lagrangian of tachyon field has the
non-canonical one.

As soon as the analysis of dynamics of expansion of the Universe \cite{sergijenko2008} doesn't allow us to choose the most preferable by the observational data model of scalar field dark energy, here we focus on study of the evolution of scalar perturbations and the large-scale structure formation in the Universe filled only with the non-relativistic matter and either classical or tachyonic field minimally coupled to it. It should be noted that the behavior of perturbations has already been studied for different classical scalar fields more widely \cite{bean2004,dave2002,unnikrishnan2008} than for tachyonic ones \cite{abramo2004,frolov2002}. The parametrizations of scalar fields, their impact on formation of the large-scale structure of the Universe as well as on cosmic microwave background anisotropy are widely discussed in the literature (see, for example, \cite{copeland2006,hu1998,hu1999,peebles2003,sahni2003} and citing therein). In this paper we analyse the models with reconstructed for $w_{de}=const$ potentials of the classical and tachyonic scalar fields and compare the obtained results to the $\Lambda$CDM-ones.

\section{Cosmological background}

We consider the homogeneous and isotropic flat Universe with metric of 4-space 
$$ds^2=g_{ij} dx^i dx^j =a^2(\eta)\left(d\eta^2-\delta_{\alpha\beta} dx^{\alpha}dx^{\beta}\right),$$
where the factor $a(\eta)$ is the scale factor, normalized  to 1 at  the current epoch $\eta_0$, $\eta$ is conformal time ($cdt=a(\eta)d\eta$). Here and below we put $c=1$, so the time variable $t\equiv x_0$  has the dimension of a length, and the latin indices $i,\,j,\,...$ run from 0 to 3, the greek ones -- over the spatial part of the metric: $\nu,\, \mu,\,...$=1, 2, 3. 

If the Universe is filled with non-relativistic matter (cold dark matter and baryons) and minimally coupled dark energy, the dynamics of its expansion is completely described by the Einstein equations 
\begin{eqnarray}
\mathcal{R}_{ij}-{\frac{1}{2}}g_{ij}\mathcal{R}=8\pi G \left(\mathcal{T}_{ij}^{(m)}+\mathcal{T}_{ij}^{(de)}\right),
\label{Einstein-eq}
\end{eqnarray}
where $\mathcal{R}_{ij}$ is the Ricci tensor and $\mathcal{T}_{ij}^{(m)}$, $\mathcal{T}_{ij}^{(de)}$ -- energy-momentum tensors of matter $(m)$ and dark energy $(de)$. If these components interact only gravitationally then each of them satisfy differential energy-momentum conservation law separately:
\begin{eqnarray}
\mathcal{T}^{i\;\;(m,de)}_{j\;;i}=0
\label{conserv-eq}
\end{eqnarray}
(here and below ``;{\it i}'' denotes the
covariant derivative with respect to the coordinate $x^i$). For the perfect fluid with density $\rho_{(m,de)}$ and pressure $p_{(m,de)}$, related by the equation of state $p_{(m,de)}=w_{(m,de)}\rho_{(m,de)}$, it gives
\begin{eqnarray}
\dot{\rho}_{(m,de)}=-3\frac{\dot a}{a} \rho_{(m,de)}(1+w_{(m,de)}) \label{eqconsm}
\end{eqnarray}
(here and below a dot over the variable denotes the derivative with respect to the conformal time: ``$\dot{\;\;}$''$\equiv d/d\eta$). The matter is considered to be non-relativistic, so  $w_m=0$ and $\rho_m=\rho_m^{(0)}a^{-3}$ (here and below ``0'' denotes the present values). 

We assume the dark energy to be a scalar field with either classical Lagrangian 
\begin{eqnarray}
\mathcal{L}_{clas}=\frac{1}{2}\phi_{;i}\phi^{;i}-\mathcal{U}(\phi) \label{lagr_cf}
\end{eqnarray}
or Dirac-Born-Infeld (tachyonic) one
\begin{eqnarray}
 \mathcal{L}_{tach}=-\mathcal{U}(\xi)\sqrt{1-\xi_{;i}\xi^{;i}},
\end{eqnarray}
where $\phi$, $\xi$ are the classical and tachyonic fields respectively while $\mathcal{U}(\phi)$, $\mathcal{U}(\xi)$ are the field potentials defining the models.
We suppose also the background scalar fields to be homogeneous, so their energy densities and pressures depend only on time:
\begin{eqnarray}
&&\rho_{clas}=\frac{1}{2a^2}\dot{\phi}^{2}+\mathcal{U}(\phi),\,\,\,\,\,\,\,\,
p_{clas}=\frac{1}{2a^2}\dot{\phi}^{2}-\mathcal{U}(\phi),\\
&&\rho_{tach}=\frac{\mathcal{U}(\xi)}{\sqrt{1-\dot{\xi}^2/a^2}},\,\,\,\,\,\,\,\,\,\,\,
p_{tach}=-\mathcal{U}(\xi)\sqrt{1-\frac{\dot{\xi}^2}{a^2}}.
\end{eqnarray}
Then the conservation law gives the scalar field evolution equations
\begin{eqnarray}
 &&\ddot{\phi}+2aH\dot{\phi}+a^2\frac{d\mathcal{U}}{d\phi}=0, \\
&&\frac{\ddot{\xi}-aH\dot{\xi}}{1-\left(\dot{\xi}/a\right)^2}+3aH\dot{\xi}+\frac{a^2}{\mathcal{U}}\frac{d\mathcal{U}}{d\xi}=0,
\end{eqnarray}
where $H=\dot{a}/{a^2}$ is the Hubble constant for any moment of the conformal time $\eta$.

We specify the model of each field using the EoS parameter $w_{de}\equiv p_{de}/\rho_{de}$. It is obvious that the scalar field evolution equations have the analytical solutions for $w=const$ (here and below we omit index $de$ denoting both -- classical and tachyonic -- scalar fields for $w_{de}$). In this case another important thermodynamical parameter -- the adiabatic sound speed $c^2_a\equiv \dot{p}_{de}/\dot{\rho}_{de}$ -- is equal to $w$.

The analysis of the dynamics of the Universe expansion for the reconstructed fields with $w=const$ was presented in \cite{sergijenko2008}. It doesn't depend on the scalar field Lagrangian and -- as a result -- doesn't allow us to distinguish such models of scalar fields. So, in order to choose the most adequate to observations type of dark energy we should study at least the linear stage of the evolution of scalar perturbations.

\section{Evolution of scalar linear perturbations}
We derive the equations of evolution of scalar linear perturbations in dark energy -- matter dominant era by varying of the Lagrange-Euler and Einstein equations in the conformal-Newtonian frame with space-time metric
\begin{eqnarray}
ds^2&=&a^2(\eta)[(1+2\Psi(\textbf{x},\eta))d\eta^2-(1+2\Phi(\textbf{x},\eta))\delta_{\alpha\beta}dx^{\alpha}dx^{\beta}],
\end{eqnarray}
where $\Psi(\textbf{x},\eta)$ and $\Phi(\textbf{x},\eta)$ are metric perturbations, which in the case of zero proper anisotropy of medium (as for dust matter and scalar fields) satisfy the condition $\Psi(\textbf{x},\eta)=-\Phi(\textbf{x},\eta)$ exactly \cite{bardeen1980,kodama1984}. In the theory of linear perturbations all spatially-dependent variables are usualy Fourier-transformed, so, all perturbations -- of metric, fields, matter density and velocity -- in equations are presented by their Fourier amplitudes: $\Psi(k,\eta)$), $\delta{\phi}(k,\eta)$, $\delta{\xi}(k,\eta)$,  $\delta^{(m)}(k,\eta)$, 
$V^{(m)}(k,\eta)$ etc., where  $k$ is the wave number. They are gauge-invariant -- as it is particularly discussed in the original papers \cite{bardeen1980,kodama1984} and numerous reviews (see, for example, \cite{durrer2001,copeland2006,novosyadlyj2007} and citing therein). The energy density and velocity perturbations of dark energy, $\delta^{(de)}$ and $V^{(de)}$, are connected with the perturbations of field variables $\delta{\phi}$, $\delta{\xi}$ in the following way:
\begin{eqnarray}
&&\delta^{(clas)}=(1+w)\left(\frac{\dot{\delta{\phi}}}{\dot{\phi}}-\Psi+\frac{a^2\delta{\phi}}{\dot{\phi}^2}\frac{d\mathcal{U}}{d\phi}\right),\label{dq}\\
&&V^{(clas)}=\frac{k\delta{\phi}}{\dot{\phi}},\label{vq}\\
&&\delta^{(tach)}=-\frac{1+w}{w}\left(\frac{\dot{\delta\xi}}{\dot{\xi}}-\Psi\right)+\frac{1}{\mathcal{U}}\frac{d\mathcal{U}}{d\xi}\delta{\xi},\\
&&V^{(tach)}=\frac{k\delta{\xi}}{\dot{\xi}}.\label{vt}
\end{eqnarray}

Other non-vanishing gauge-invariant perturbations of scalar field are isotropic pressure perturbations 
\begin{eqnarray}
&&\pi_L^{(clas)}=\frac{1+w}{w}\left(\frac{\dot{\delta{\phi}}}{\dot{\phi}}-\Psi-\frac{a^2\delta{\phi}}{\dot{\phi}^2}\frac{d\mathcal{U}}{d\phi}\right),\\
&&\pi_L^{(tach)}=\frac{1+w}{w}\left(\frac{\dot{\delta\xi}}{\dot{\xi}}-\Psi\right)+\frac{1}{\mathcal{U}}\frac{d\mathcal{U}}{d\xi}\delta{\xi}
\end{eqnarray}
and intrinsic entropy
\begin{eqnarray}
\Gamma^{(de)}=\pi_L^{(de)}-\frac{c^2_a}{w}\delta^{(de)}.
\end{eqnarray}
The density perturbation of any component in the conformal-Newtonian gauge $D_s\equiv\delta$, which is gauge-invariant variable,
 is related to the other gauge-invariant variables of density perturbations $D$ and $D_g$ as:
\begin{eqnarray}
  D=D_g+3(1+w)\left(\Psi+\frac{\dot{a}}{a}\frac{V}{k}\right)=D_s+3(1+w)\frac{\dot{a}}{a}\frac{V}{k},\label{ddgds}
\end{eqnarray}
where $D_s$, $D$, $D_g$ and $V$ correspond to either $m$- or $de$-component. 

\subsection{Evolution equations}

Evolution equations for scalar field perturbations $\delta{\phi}(k,\eta)$ and $\delta{\xi}(k,\eta)$ can be obtained either from Lagrange-Euler equation or from differential energy-momentum conservation law ${\mathcal{T}^i_{0;i}}^{(de)}=0$:
 \begin{eqnarray}
&&\ddot{\delta{\phi}}+2aH\dot{\delta{\phi}}+\left[k^2+a^2\frac{d^2\mathcal{U}}{d\phi^2}\right]\delta{\phi}+2a^2\frac{d\mathcal{U}}{d\phi}\Psi
-4\dot{\Psi}\dot{\phi}=0,\label{phi_evol}\\
&&\ddot{\delta{\xi}}+\left[2aH-9aH\left(\frac{\dot{\xi}}{a}\right)^2-\frac{2}{\mathcal{U}}\frac{d\mathcal{U}}{d\xi}\dot{\xi}\right]\dot{\delta{\xi}}+\left[k^2+a^2\left(\frac{1}{\mathcal{U}}\frac{d^2\mathcal{U}}{d\xi^2}-\left(\frac{1}{\mathcal{U}}\frac{d\mathcal{U}}{d\xi}\right)^2\right)\right]\times\nonumber\\
&&\left(1-\left(\frac{\dot{\xi}}{a}\right)^2\right)\delta{\xi}-\dot{\Psi}\dot{\xi}-3\dot{\Psi}\dot{\xi}\left(1-\left(\frac{\dot{\xi}}{a}\right)^2\right)+2\Psi \frac{a^2}{\mathcal{U}}\frac{d\mathcal{U}}{d\xi}+
6aH\Psi\dot{\xi}\left(\frac{\dot{\xi}}{a}\right)^2=0.
\end{eqnarray}   
The linearised Einstein equations for gauge-invariant perturbations of metric and energy-momentum tensor components are  
\begin{eqnarray}
&&\dot{\Psi}+aH\Psi-\frac{4\pi Ga^2}{k}\left[\rho_mV^{(m)}+\rho_{de}(1+w)V^{(de)}\right]=0,\\
&&\dot{V}^{(m)}+aHV^{(m)}-k\Psi=0,\\
&&\dot{D_g}^{(m)}+kV^{(m)}=0,\\
&&\dot{V}^{(de)}+aH(1-3c_a^2)V^{(de)}-k(1+3c_a^2)\Psi-\frac{c_a^2k}{1+w}D_g^{(de)}-\frac{wk}{1+w}\Gamma^{(de)}=0,\\
&&\dot{D_g}^{(de)}+3(c_a^2-w)aHD_g^{(de)}+k(1+w)V^{(de)}+3aHw\Gamma^{(de)}=0,
\end{eqnarray}
where
\begin{eqnarray}
 w\Gamma^{(clas)}=(1-c_a^2)\left[D_g^{(clas)}+3(1+w)\Psi+3aH(1+w)\frac{V^{(clas)}}{k}\right]=(1-c_a^2)D^{(clas)},&&\\
 w\Gamma^{(tach)}=-(w+c_a^2)\left[D_g^{(tach)}+3(1+w)\Psi+3aH(1+w)\frac{V^{(tach)}}{k}\right]=-(w+c_a^2)D^{(tach)}.&&
\end{eqnarray}
In $w=const$-case $1-c_a^2=1-w$, $w+c_a^2=2w$, hence the difference between equations for classical and tachyonic fields isn't big (for $w$ close to $-1$ -- as it follows from the observable data \cite{komatsu2008}) and suggests the similarity of their solutions.

So, in each case we have the system of 5 first-order ordinary differential equations for 5 unknown functions $\Psi(k,a)$, $D_g^{(m)}(k,a)$, $V^{(m)}(k,a)$, $D_g^{(de)}(k,a)$ and $V^{(de)}(k,a)$ satisfying also the constraint equation:
\begin{eqnarray}
 -k^2\Psi=4\pi Ga^2\left(\rho_mD^{(m)}+\rho_{de}D^{(de)}\right).\label{constrainteq}
\end{eqnarray}

\subsection{Initial conditions}

Now we are going to specify the adiabatic initial conditions. The adiabaticity condition in two-component model gives $D_g^{(m)}=D_g^{(de)}/(1+w)$ \cite{doran2003,durrer2001,kodama1984}.

Since the density of the $w=const$-fields is negligible at the early epoch ($a\ll 1$), both our models are initially matter-dominated. It is known that in such case the growing mode corresponds to $\Psi=const$. The field equations of motion for the reconstructed potentials are following:
\begin{eqnarray}
&&\delta\phi''+\left(\frac{5}{2}-\frac{3w}{2}\frac{\Omega_{de}a^{-3w}}{1-\Omega_{de}+\Omega_{de}a^{-3w}}\right)\frac{\delta\phi'}{a}+\left[\frac{k^2}{H_0^2a(1-\Omega_{de}+\Omega_{de}a^{-3w})}+\right. \nonumber\\
&&\left.\frac{9(1-w)}{4a^2}\left(2+w+w\frac{\Omega_{de}a^{-3w}}{1-\Omega_{de}+\Omega_{de}a^{-3w}}\right)\right]\delta\phi- \nonumber\\
&&a^{-\frac{3w}{2}}\sqrt{\frac{3}{8\pi G}\frac{\Omega_{de}(1+w)}{1-\Omega_{de}+\Omega_{de}a^{-3w}}}\frac{4a\Psi'+3(1-w)\Psi}{a^2}=0 
\end{eqnarray}
for the classical field and
\begin{eqnarray}
 &&\delta\xi''-\left(\frac{1}{2}+3w+\frac{3w}{2}\frac{\Omega_{de}a^{-3w}}{1-\Omega_{de}+\Omega_{de}a^{-3w}}\right)\frac{\delta\xi'}{a}-w\left[\frac{k^2}{H_0^2a(1-\Omega_{de}+\Omega_{de}a^{-3w})}+\right. \nonumber\\
&&\left.\frac{9}{2a^2}\left(1+w\frac{\Omega_{de}a^{-3w}}{1-\Omega_{de}+\Omega_{de}a^{-3w}}\right)\right]\delta\xi- \nonumber\\
&&\frac{\sqrt{1+w}}{H_0\sqrt{1-\Omega_{de}+\Omega_{de}a^{-3w}}}\frac{(1-3w)a\Psi'-6w\Psi}{\sqrt{a}}=0 
\end{eqnarray}
for the tachyonic one. Here and below a prime denotes the derivative with respect to the scale factor $a$ and $\Omega_{de}=\rho_{de}/\rho_c$, where $\rho_c\equiv3H_0^2/8\pi G$.

The condition $\Psi=const$ for $a\ll 1$ gives:
\begin{eqnarray}
 &&\delta{\phi}=\frac{1}{\sqrt{6\pi G}}\sqrt{\frac{\Omega_{de}(1+w)}{1-\Omega_{de}}}\Psi a^{-\frac{3w}{2}},\\
&&\delta{\xi}=\frac{2}{3}\frac{\sqrt{1+w}}{H_0\sqrt{1-\Omega_{de}}}\Psi a^{\frac{3}{2}}.
\end{eqnarray}
Here $\Gamma^{(de)}=0$. 

Using these solutions and equations (\ref{dq})-(\ref{vt}), (\ref{ddgds}), (\ref{constrainteq}), one can find the initial values of $D_g^{(m)}(k,a)$, $V^{(m)}(k,a)$, $D_g^{(de)}(k,a)$, $V^{(de)}(k,a)$:
\begin{eqnarray}
&&{V^{(de)}}_{init}=\frac{2}{3}\frac{k}{H_0}\frac{\Psi_{init}}{\sqrt{1-\Omega_{de}}}\sqrt{a_{init}},\\
&&{D_g^{(de)}}_{init}=-5(1+w)\Psi_{init},\\
&&{V^{(m)}}_{init}=\frac{2}{3}\frac{k}{H_0}\frac{\Psi_{init}}{\sqrt{1-\Omega_{de}}}\sqrt{a_{init}},\\
&&{D_g^{(m)}}_{init}=-5\Psi_{init},
\end{eqnarray}
which specify the growing mode of the adiabatic perturbations.

\subsection{Numerical analysis}
We have integrated numerically the systems of equations for dust matter and dark energy with $w=const$ for the adiabatic initial conditions using the publicly available code DVERK\footnote[1]{It is created by T.E. Hull, W.H.Enright, K.R. Jackson in 1976 and is available at  http://www.cs.toronto.edu/NA/dverk.f.gz}. We used the set of cosmological parameters from {\it http://lambda.gsfc.nasa.gov/product/map}, assumed $\Psi_{init}=-1$, $a_{init}=10^{-10}$ and integrated up to $a=1$. The evolution of perturbations is scale dependent, so we performed calculations for $k=0.0001,\,\,0.001,\,\,0.01$ and $0.1$ Mpc$^{-1}$. The models with the classical scalar field are denoted as QCDM, with tachyon -- as TCDM. For comparison we also solve the evolution equations for the $\Lambda$CDM-model.

\begin{figure}[ht]
\centerline{\includegraphics[height=6cm]{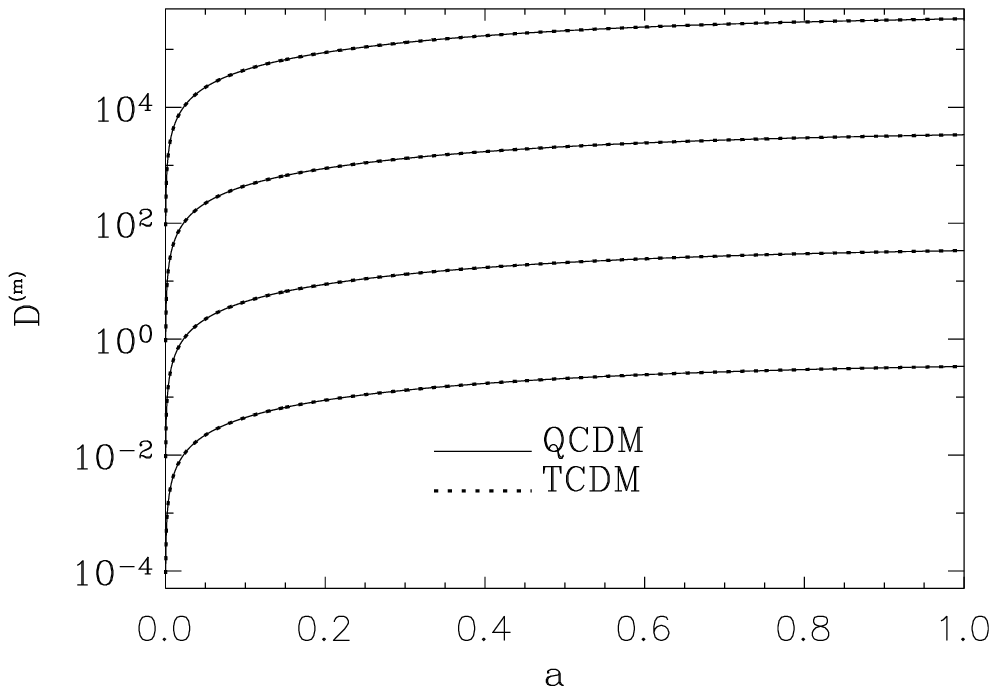}
\includegraphics[height=6cm]{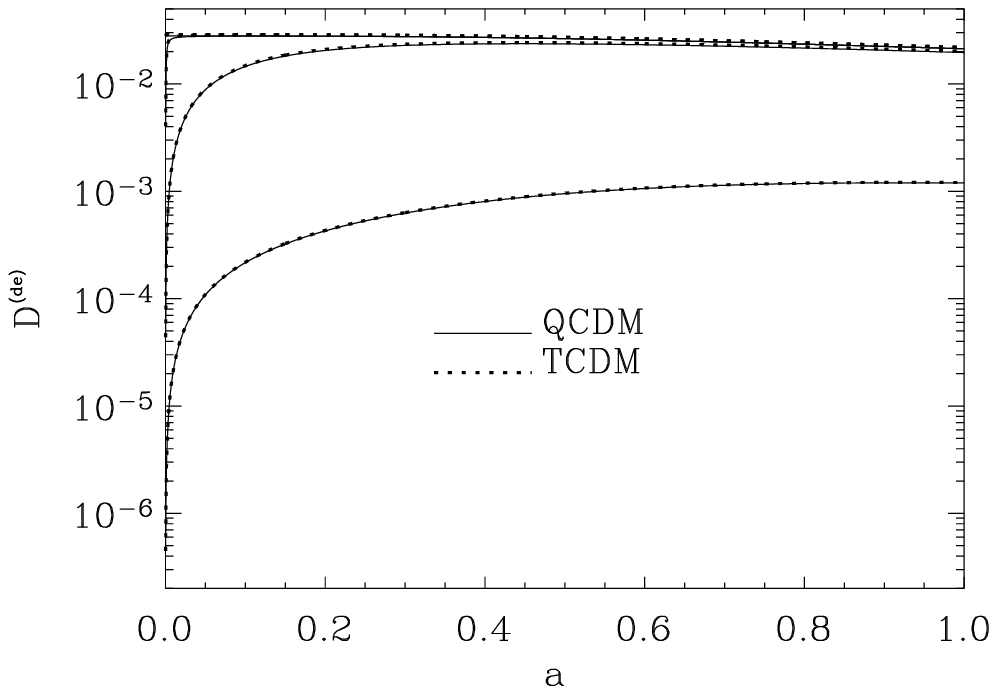}}
\centerline{\includegraphics[height=6cm]{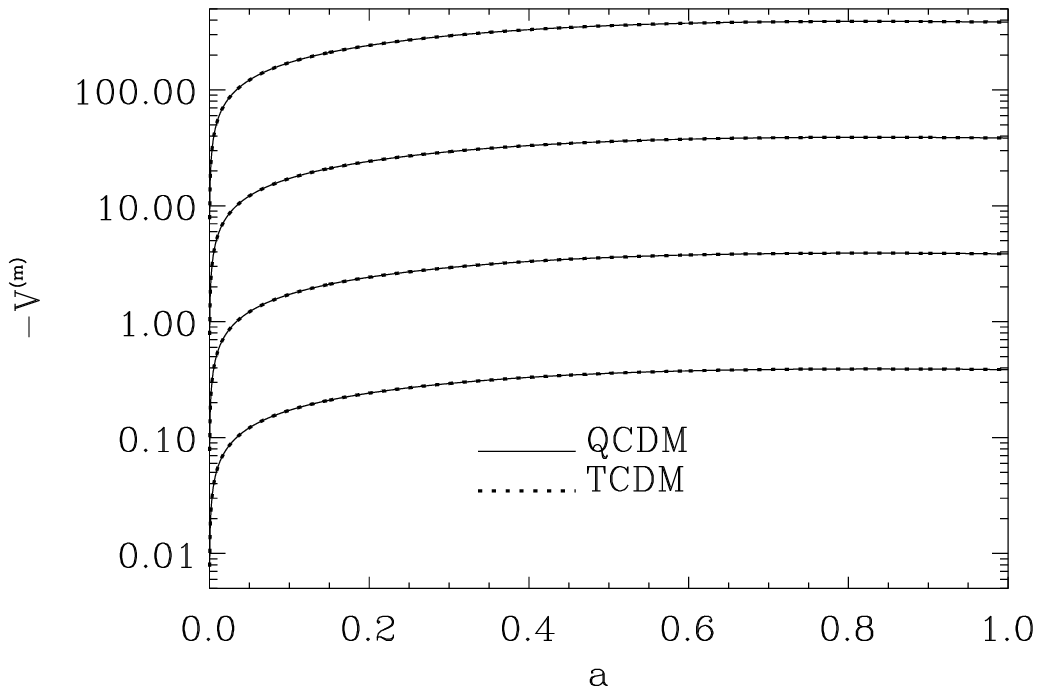}
\includegraphics[height=6cm]{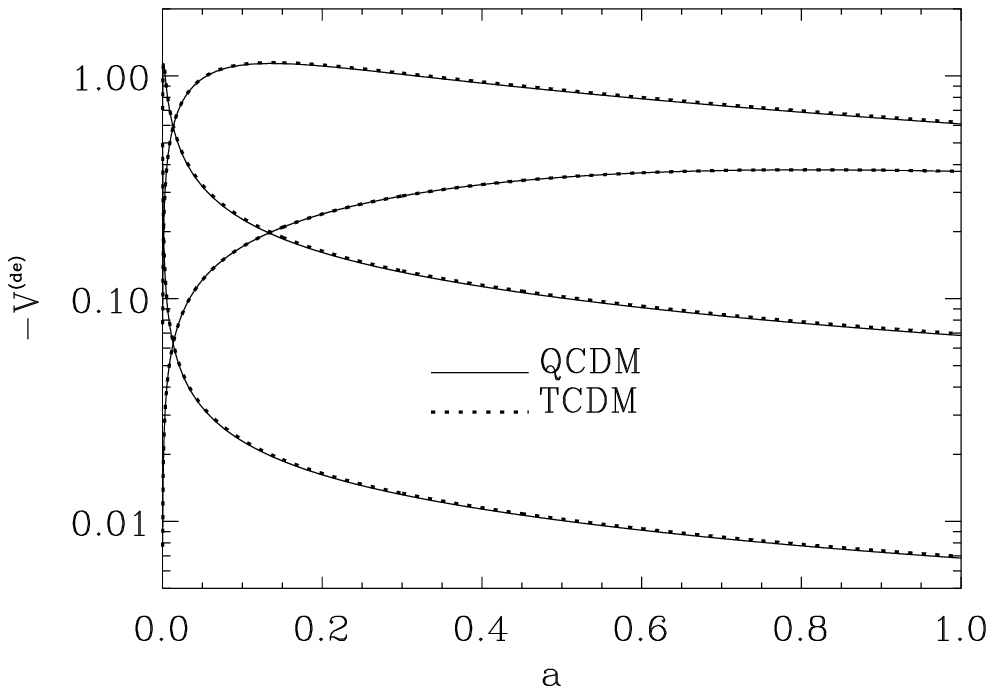}}
\caption{Evolution of the density (top) and velocity (bottom) perturbations for the non-relativistic matter (left) and dark energy (right). In the left column the scales are $k=0.1$, $0.01$, $0.001$ and $0.0001$ $Mpc^{-1}$ from top to bottom. In the right column they are also $k=0.1$, $0.01$, $0.001$ and $0.0001$ $Mpc^{-1}$ from top to bottom for the density perturbations while for the velocity ones the curves correspond to $k=0.001$, $0.0001$, $0.01$ and $0.1$ $Mpc^{-1}$ from top to bottom at $a\approx1$. The cosmological parameters are: $\Omega_{de}=0.722$, $w=-0.972$, $\Omega_m=0.278$, $h=0.697$.}
\label{pg}
\end{figure}

As it can be seen in Fig.\ref{pg}, the simple conclusion, that the behavior of the scalar linear perturbations in the model with the tachyonic field with $w=const$ should be similar to that in the model with the corresponding classical field, is valid. The curves for both fields almost overlap, so in this case it is not possible to choose the Lagrangian prefered by observations (see also \cite{unn2008}).

In both models studied here the matter clusters while dark energy is smoothed out on subhorizon scales. Generally, at present epoch the growth of the matter density perturbations is supressed and -- unlike $\Lambda$CDM-case -- such supression is scale dependent, however this dependence is very weak. The dark energy perturbations grow approximately up to the moment of the entering of particle horizon and start to decay after that (the density perturbation $D^{(de)}$ -- slowly).

Note that the perturbations in such scalar field models are insensitive to the initial conditions. Really, if we assume the dark energy to be initially homogeneous ($\delta\phi=\delta\phi'=0$ and $\delta\xi=\delta\xi'=0$), the results of numerical integration will be the same as in adiabatic case (the similar conclusion was made in \cite{dave2002,liu2004}).

\begin{figure}
\centerline{
\includegraphics[width=10cm]{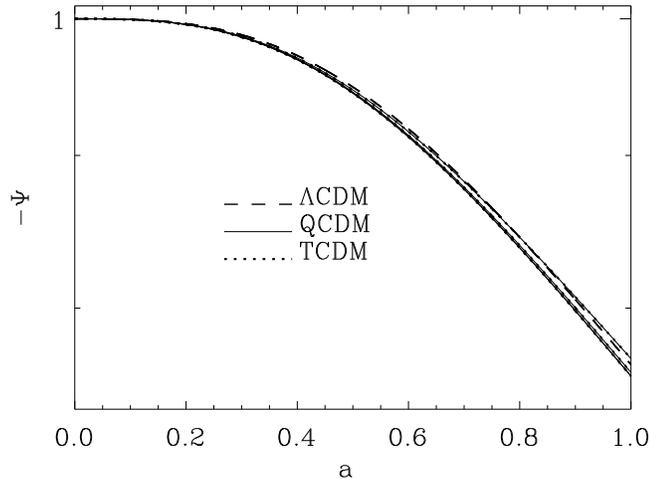}}
\caption{Evolution of the gravitational potential for the scales $k=0.0001$, $0.001$, $0.01$ and $0.1$ $Mpc^{-1}$ (from top to bottom). The curves for QCDM- (solid line) and TCDM- (dotted) models with the dark energy perturbations overlap.}
\label{psi}
\end{figure}

The simplest test for identification of the source causing the accelerated expansion of the Universe can be based on the action of the studied fields on cosmic microwave background. Here the main attention should be paid to the temporal variation of the gravitational potential which causes the late-time integrated Sachs-Wolfe (ISW) effect. In Fig.\ref{psi} the evolution of $\Psi$ is shown for both scalar field models and for $\Lambda$CDM-one for the scales of perturbations $k=0.0001$, $0.001$, $0.01$ and $0.1$ $Mpc^{-1}$. It can be easily seen that the scale dependence is weak and there is no substantial difference between classical and tachyonic dark energy. Such scalar fields are in many senses similar to the cosmological constant and their behavior is closer to that of the $\Lambda$-term for EoS parameter values closer to $-1$. The given dependence for $\Lambda$CDM-model doesn't allow us to exclude this model using the observational data, because the difference between it and those in models with the scalar field dark energy is not substantial.

It should be noted that neglecting of the dark energy perturbations leads to the ``quasi-$\Lambda$CDM''-models, i.e. models in which the fields affect the growth of the matter perturbations only through the background. In these models the decay of the gravitational potential is scale independent and close to the small-scale one in the models with perturbed dark energy (in agreement with the results of \cite{liu2004}). 

Another possible test is based on the study of action of dark energy on the clustering properties of dust matter. However, here we need the analysis of the evolution of scalar perturbations at the non-linear stage.

\section{Spherical collapse in the models with homogeneous dark energy}

The simplest and most popular approach used in the study of the non-linear stage of the large-scale structure formation is the spherical collapse model. Within this framework we analyse the formation of the virialised haloes in the $\Lambda$CDM- and in the $w=const$ QCDM- and TCDM-models with reconstructed potentials, discussed in the previous sections.

The magnitudes of density perturbations of the classical and tachyonic scalar fields with scale less than the particle horizon are lower than corresponding magnitudes of the matter density ones by few orders and practically do not affect their growth. The amplitudes of matter density perturbations in the QCDM- and TCDM-models grow almost equally in cosmologies with the same parameters. They are also close to ones in the $\Lambda$CDM-models. So, in order to simplify the discussion of the non-linear evolution of scalar perturbations we assume the dark energy component to tend to homogeneity. Other reasons for homogeneous distribution of dark energy in the regions of matter inhomogeneties see, for example, in \cite{Maor05,Mota04}. Hence, the temporal dependence of the dark energy density is defined by the corresponding background equation. (Structure formation in inhomogeneous dark energy models has been analysed in \cite{nunes2006}.)

The relative perturbation of mass of the dust component in the comoving volume $v=4\pi R^3/3$ and metric $ds^2 = dt^2 - \mathcal{M}^2(R)y^2(t,R)dR^2 -x^2(t,R)R^2(\cos^2\theta d\varphi^2+d\theta^2)$ is following \cite{kul08}:
\begin{equation}
 \delta_m = \left(\frac{a(t)}{x(t,R)}\right)^3 - 1,
\end{equation}
where $x(t,R)$ is the local scale factor derived from the Einstein equation $\mathcal{G}^1_1 = \mathcal{\kappa} \mathcal{T}^1_1$ (here $\mathcal{G}_j^i$ is the Einstein tensor and $\mathcal{\kappa}$ is the Einstein constant) \cite{kul08,kul07}:
\begin{equation}
 \ddot x = -\frac32 \frac{p_{de}}{\rho_c}x-\frac12 \frac{{\dot x}^2}{x} + \frac{1}{x}\frac{\Omega_f}{2}\label{ddx}
\end{equation}
(in this section ``$\dot{\;\;}$''$\equiv d/H_0dt$).

For the $\Lambda$-term we should put $p_{de}/\rho_c = - \Omega_{\Lambda}$ while for the quintessential dark energy the relation is $p_{de}/\rho_c = w\Omega_{de}$. The local curvature parameter $\Omega_f$ gives the amplitude of the initial perturbation: $\delta_m(t) \simeq \frac35(\Omega_K-\Omega_f)\Omega_m^{-1}a(t)$ at $a\ll 1$. Since in the $\Lambda$-case $\Omega_f\equiv \Omega_f(R)$ for the dark energy we should put $\Omega_f\equiv \Omega_f(t,R)$ \cite{wang98,kul07}. It means that here we need an additional equation defining the evolution of the local curvature. However, for the homogeneous perturbations ($\frac{\partial}{\partial R}\Omega_f = 0$, $x\equiv x(t) = y(t)$) using the combination of the Einstein equations $\mathcal{G}^0_0 + \mathcal{G}^1_1 + 2\mathcal{G}^2_2 = \mathcal{\kappa} (\mathcal{T}^0_0 + \mathcal{T}^1_1 + 2\mathcal{T}^2_2)$ we obtain the motion equation without time-dependent curvature \cite{wang98}:
\begin{equation}
 \frac{\ddot x}{x} = - \frac{1}{\rho_c}\left(3p_{de}+\rho_{de} + \rho_m\right),\label{ws98}
\end{equation}
where the dust matter density is $\rho_m=\rho_m^{(0)} x^{-3}$. 
Combining of the equation (\ref{ws98}) and Friedmann equations for the homogeneous Universe allows us to find the evolution of the mass perturbation. 

In the analysis of halo formation the moment of reaching of the dynamical equilibrium is important. According to the virial theorem at this moment the kinetic energy becomes $T_{vir}=-\frac12U_{m,vir} + U_{\Lambda,vir}$.
In the $\Lambda$CDM-model the energy conservation law $T+U_m+U_{\Lambda}=E$ is obtained by integration of the equation of motion (\ref{ddx}), multiplied by $x\dot x$. From this follows: $T=\frac12{\dot x}^2$, $U_m = -\frac12\Omega_mx^{-1}$, $U_{\Lambda} = -\frac12 \Omega_{\Lambda}x^2$ and $E = \frac12 \Omega_f$. In models with the $\Lambda$-term the total energy is constant in time and at the reaching of the dynamical equilibrium is equal $E = \frac12U_{m,vir} + 2U_{\Lambda,vir}$. Alternatively, at the turnaround moment, when $\dot x = T = 0$, the total energy is $E = U_{m,ta} + U_{\Lambda,ta}$. From these equations we obtain finally: $\frac12U_{m,vir} + 2U_{\Lambda,vir} = U_{m,ta} + U_{\Lambda,ta}$. This identity, valid for  the $\Lambda$-case, doesn't hold for the dark energy, since the temporal variation of the curvature in the perturbed region gives $E(t_{vir})\ne E(t_{ta})$.
In other words, explicit temporal dependence of the dark energy density $\rho_{de}(t)$ leads to the explicit dependence of the potential energy of this component $U_{de}(t,x)$ on time and -- hence -- to the explicit temporal dependence of the total energy: $E(t) = T(\dot x) + U_{m}(x) + U_{de}(t,x)$. It means that at different times we have different values of the total energy. For estimation of the moment of reaching of the dynamical equilibrium for the dark energy case we use the equations (3.13)-(3.17) from \cite{kul07}. These equations describe the evolution of the spherically-symmetric perturbation with the arbitrary profile in the model with dark energy. Assuming there $\Omega_f = \Omega_f(t)$,
$x\equiv x(t) = y(t)$ and $V\equiv V(t)$, we obtain the equations for the homogeneous spherical cloud. The additional condition of homogeneity of the dark energy (the equality of (\ref{eqconsm}) and (3.17) from \cite{kul07}) gives the expression for $V$. Using it together with (3.15) from \cite{kul07} we obtain the equation describing the temporal variation of curvature:
\begin{equation}
 \dot\Omega_f = 3\left(\frac{\dot a}{a}-\frac{\dot x}{x}\right)(1+w)\Omega_{de}x^2.\label{dof}
\end{equation}
Combining of this equation with (\ref{ddx}) leads to the energy conservation equation $E(t) = T(\dot x) + U_{m}(x) + U_{de}(t,x)$ in the following form:
\begin{equation}
 \frac{\dot x^2}{2} -\frac12\frac{\Omega_m}{x}-\frac12 \Omega_{de}x^2 = \frac12\Omega_f.\label{eq}
\end{equation}
Really, it can be easily seen that differentiating (\ref{eq}) with respect to time and using (\ref{eqconsm}), (\ref{dof}) we obtain (\ref{ws98}). Combining it with (\ref{eq}) we get (\ref{ddx}). Since $E=\frac12\Omega_f$, the equation (\ref{dof}) describes the temporal variation of the total energy. Using the virial theorem, for the moment of reaching of  the dynamic equilibrium we write: $E(t_{vir}) = \frac12U_{m,vir} + 2U_{de,vir}$. At the turnaround moment we have: $E(t_{ta}) = U_{m,ta} + U_{de,ta}$.  Taking into account that $E(t_{vir})\ne E(t_{ta})$, we obtain finally:
 \begin{equation}
 \frac12U_{m,vir} + 2U_{de,vir} = \frac{E(t_{vir})}{E(t_{ta})}\left(U_{m,ta} + U_{de,ta}\right).
\end{equation}
This equality differs from used one in \cite{lokas02,wang98} by factor $E(t_{vir})/E(t_{ta})$, which, however, is close to unity.
\begin{figure}[htbp]
\centerline{\includegraphics[width=10.0cm]{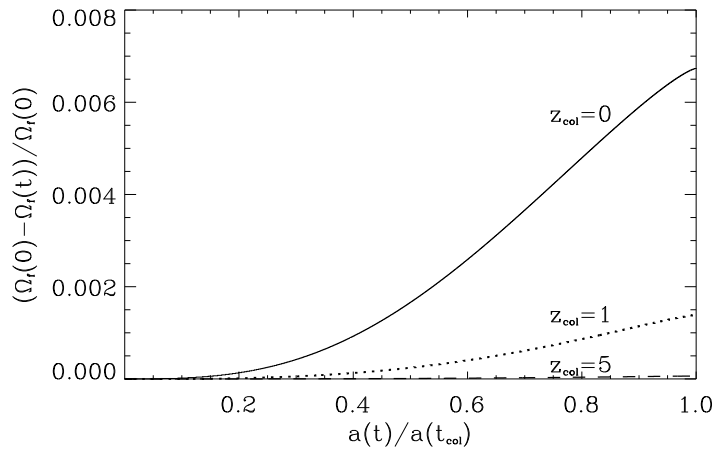}}
\caption{
The temporal dependences of the local curvature parameter change  $\left(\Omega_f(0)-\Omega_f(t)\right)/\Omega_f(0)$ for 3 different $\Omega_f(0)$, which correspond to the redshifts of halo collapse $z_{col}=0,\,1,\,5$.}
\label{of}
\end{figure}

The mentioned above density contrast $\Delta_c$ is defined as the ratio of the density of virialised halo to the critical one at the expected moment of collapse: $\Delta_c = \rho_{vir}/\rho_c(t_{col})= \Omega_mx_{vir}^{-3}(H_0/H(t_{col}))^2$. So, using the energy conservation law we find:
\begin{equation}
 4\Omega_{de}(t_{col})x_{vir}^3+2\Omega_f(t_{col})x_{vir}+\Omega_m=0.
\end{equation}
We choose the initial value of the curvature in the perturbed region $\Omega_f(0)$ by setting the moment of  collapse $t_{col}$.
Unfortunately, in the scalar field plus CDM model the equation (\ref{eq}) isn't symmetrical in time with respect to the moment of turnaround as it was for the $\Lambda$CDM-model.
Hence the relation between the quantities $t_{col}$ and $\Omega_f(0)$ must be established by the numerical integration of the equations (\ref{dof})-(\ref{eq}).
It is interesting to see how the curvature changes in the models with dark energy plus CDM, since in the $\Lambda$CDM-models it remains constant.
In Fig.\ref{of} we show such dependences for 3 different initial values of the local curvature parameter, which correspond to the redshifts of halo collapse $z_{col}=0,\,1,\,5$.

\begin{figure}[htbp]
\centerline{\includegraphics[width=10.0cm]{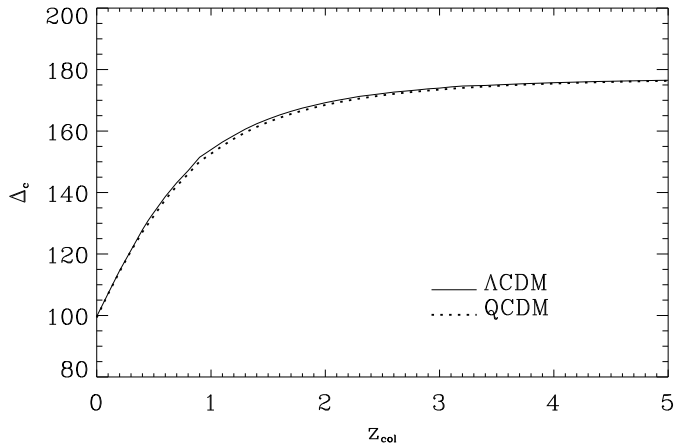}}
\caption{The density contrast $\Delta_c$ at the collapse moment $z_{col}$ of spherical cloud in the Q(T)CDM- and $\Lambda$CDM-models.}
\label{Dc}
\end{figure}
The calculations of $\Delta_c$ at the collapse moment $z_{col}$ in the $\Lambda$CDM-model ($\Omega_{\Lambda} = 0.721$, $\Omega_m = 0.279$) and in scalar field plus CDM model ($\Omega_{de} = 0.722$, $w = - 0.972$, $\Omega_m = 0.278$) are presented in Fig.\ref{Dc}. We see that the difference between the values of $\Delta_c$ in these models is insignificant for all $z_{col}$. It means that dynamics of cluster formation at all stages of their evolution -- from linear stage through collapse up to virialization -- is similar in the models with both reconstructed quintessential scalar fields if they are minimally coupled and have the same density ($\Omega_{de}$) and EoS ($w$) parameters. Practically, they are indistinguishable also from the best-fit $\Lambda$CDM-model, so, we conclude that these classes of dark energy models are degenerate with respect to their impact on dynamics of expansion of the Universe as well as formation of its large-scale structure. The other scalar field models (with special potentials, variable EoS parameter, non-minimal coupling, spatial inhomogeneity etc.) show similar but not so strongly degenerate impact on the linear and non-linear stages of structure formation (see, for example, \cite{Mota04,nunes2006}).

\section{Theoretical predictions and observations}

In this section we compare predictions of the  models with modern observational data on large-scale structure of the Universe.

We have reconstructed the potentials of classical and tachyonic scalar fields for the model with parameters $\Omega_{de}=0.722$, $w=-0.972$, $\Omega_m=0.278$, $\Omega_{b}=0.0467$, $h=0.697$, $\sigma_8=0.799$, $n_s=0.962$, taken from {\it http://lambda.gsfc.nasa.gov/product/map} 
(see also \cite{komatsu2008}). For computation of the CMB temperature and  matter density fluctuations power spectra for TCDM-model we have modified the CMBFAST-4.5.1 code substituting the subroutines evaluating the classical field perturbations by corresponding tachyonic ones. Here we used the homogeneous initial conditions for dark energy.

\begin{figure}
\centerline{
\includegraphics[height=6cm]{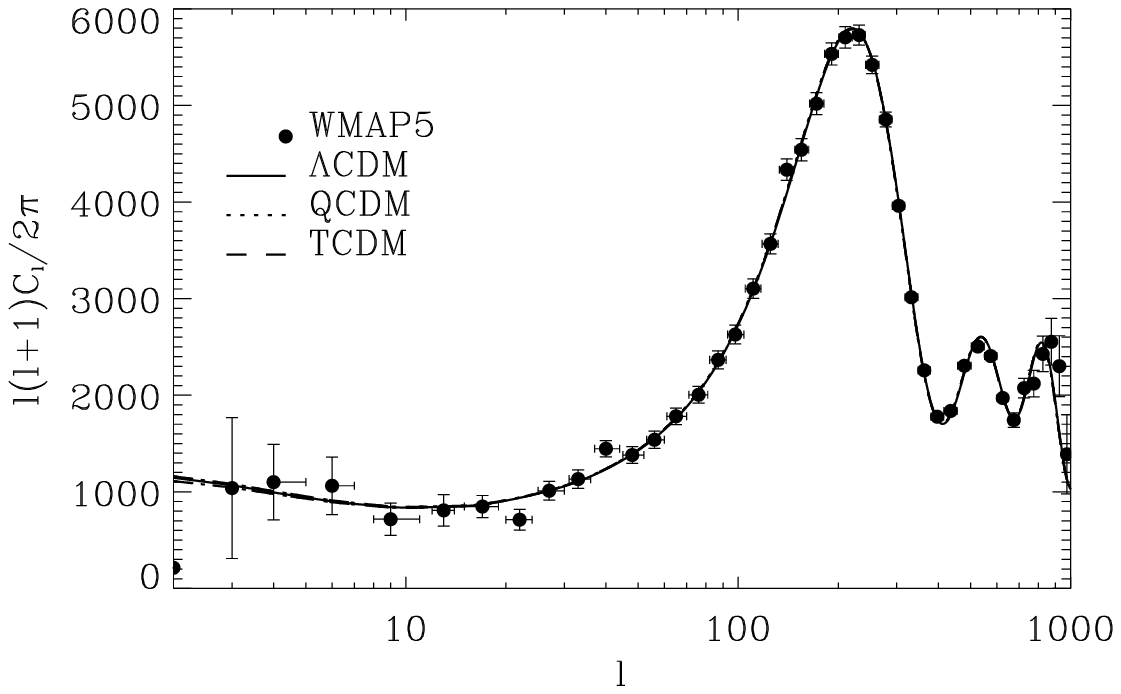}
\includegraphics[height=6cm]{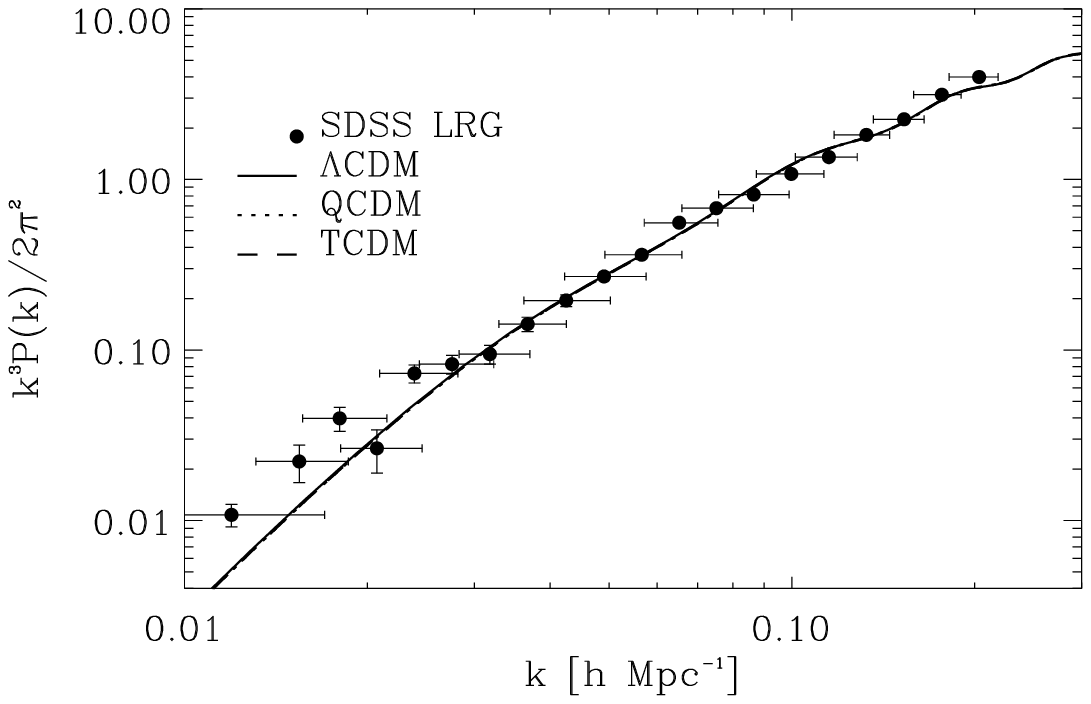}}
\caption{The CMB temperature (left) and matter density (right) fluctuations power spectra for the models of Universe with dark energy: $\Lambda$CDM -- solid line, QCDM -- dashed, TCDM -- dotted. The curves for perturbed and unperturbed dark energy (QCDM and TCDM) overlap with practically the same best-fit bias parameter ($b=1.24$). The observational CMB temperature and matter density fluctuations power spectra were obtained in the WMAP \cite{nolta2008} and SDSS \cite{tegmark2006} projects.  The amplitudes of the matter density fluctuations power spectra are normalised to WMAP 5-year data. }
\label{cl_dk}
\end{figure} 

The angular power spectra of CMB temperature fluctuations computed for QCDM- and TCDM-models are presented in the left panel of Fig.\ref{cl_dk}. There are overlaped curves for perturbed and unperturbed dark energy. For comparison the angular power spectrum of CMB temperature fluctuations for $\Lambda$CDM-model with parameters $\Omega_{de}=0.721$, $\Omega_m=0.279$, $\Omega_{b}=0.0462$, $h=0.701$, $\sigma_8=0.817$, $n_s=0.96$ \cite{komatsu2008} is computed and presented in the same figure. The corresponding five-year WMAP observational data \cite{hinshaw2008,nolta2008} are shown there too. We have renormalized the computed power spectra by fitting them to all experimental points by $\chi^2$-minimization procedure. The minimal $\chi^2$  in the $\Lambda$CDM-model equals 45.7, in the perturbed QCDM- and TCDM-models it equals 44.1 and in unperturbed ones 43.9.
So, the difference between them for 43 experimental points is statistically insignificant.

In the right panel of Fig.\ref{cl_dk} the power spectra of matter density perturbations \linebreak $P(k)\equiv<D^{(m)}\cdot D^{(m)}>$ for the same models are shown. They are normalised to 5-year WMAP data in the way explained above. The power spetrum obtained from the analysis of the clustering of the luminous red galaxies in the Sloan Digital Sky Survey (SDSS LRG) \cite{tegmark2006} is shown by dots. The model spectra we fitted to observational one using scale independent bias parameter $b$  ($P^{obs}(k)=b^2P(k)$) by the $\chi^2$-minimization procedure. The differences between $\chi^2_{min}$ of all 5 models are less than 1\%, the best-fit bias parameter $b$ equals 1.22 for $\Lambda$CDM-model and 1.24 for all scalar field plus CDM models analysed here. 

Thus, the $\Lambda$CDM-model as well as the perturbed and unperturbed QCDM- and TCDM-models can't be distinguished by current cosmological observational data.

\section*{Conclusion}
The evolution of the scalar perturbations is studied for the 2-component (dust matter and minimally coupled dark energy) cosmological models at the linear and non-linear stages. The dark energy component is assumed to be either classical or tachyonic scalar field with the potential reconstruted for the constant EoS parameter.

The evolution of linear perturbations is similar for both types of Lagrangian. The small difference can be due to the generation of the intrinsic entropy of the fields, but in $w=const$-case it is almost the same for both types of dark energy as soon as the observational data prefer the values of the EoS parameter close to $-1$.

The scalar fields studied here suppress the growth of matter density perturbations and the magnitude of gravitational potential. In these models -- unlike $\Lambda$CDM ones -- such suppression is weakly scale dependent and doesn't depend substantially on the Lagrangian.
Such features can be used for calculations of the matter density power spectrum at different redshifts and of the power spectrum of CMB temperature fluctuations in the range of scale of the later integrated Sachs-Wolfe effect and -- as a result -- for interpretation of the observable data in order to identificate the nature of dark energy. However, the higher precision of the planned experiments could probably verify only whether the EoS parameter is $-1$ ($\Lambda$CDM-model) or not, but shouldn't remove the degeneracy of the dark energy models due to the type of Lagrangian for $w=const$.

\section*{Acknowledgments}

This work was supported by the project of Ministry of Education and Science of Ukraine ``The linear and non-linear stages of evolution of the cosmological perturbations in models of the multicomponent Universe with dark energy'' (state registration number 0107U002062) and the research program of National Academy of Sciences of Ukraine ``The exploration of the structure and components of the Universe, hidden mass and dark energy (Cosmomicrophysics)'' (state registration number 0107U007279).


\begin{thebibliography}{99}
\bibitem{abramo2004} {\it Abramo L. R., Finelli F., Pereira T. S.} Constraining Born-Infeld models of dark energy with CMB anisotropies. //Phys. Rev. D.--2004.--V. 70.--id. 063517.
\bibitem{apunevych2007} {\it Apunevych S., Venhlyovska B., Kulinich Yu., Novosyadlyj B.}  WMAP-2006: cosmological parameters and the large scale structure of the Universe. // Kinematics and Physics of Celestial Bodies.--2007.--V. 23, No 2.--P. 67--82.
\bibitem{bagla2003}{\it Bagla J. S., Jassal H. K., Padmanabhan T.} Cosmology with tachyon field as dark energy // Phys. Rev. D.--2003.--V. 67.--id. 063504.
\bibitem{bardeen1980}{\it Bardeen J. M.} Gauge-invariant cosmological perturbations // Phys. Rev. D. -- 1980. -- V. 22. -- P. 1882--1905. 
\bibitem{bean2004}{\it Bean R., Dore O.} Probing dark energy perturbation: The dark energy equation of state and speed of sound as measured by WMAP // Phys. Rev. D.--2004.--V. 69.--id. 083503.
\bibitem{dave2002}{\it Dave R., Caldwell R. R., Steinhardt P. J.} Sensitivity of the cosmic microwave background anisotropy to initial conditions in quintessence cosmology // Phys. Rev. D.--2002.--V. 66.--id. 023516.
\bibitem{doran2003}{\it Doran M., Muller C. M., Schafer G., Wetterich C.} Gauge-invariant initial conditions and early time perturbations in quintessence universe // Phys. Rev. D.--2004.--V. 68.--id. 063505.
\bibitem{durrer2001}{\it Durrer R.} The theory of CMB anisotropies // Journal of Physical Studies. -- 2001. -- V. 5. -- P. 177--214.
\bibitem{copeland2006} {\it Copeland E.J., Sami M., Tsujikava S.}   Dynamics of Dark Energy // Int. J. Mod. Phys. D.--2006.--V. 15.--P. 1753--1935. 
\bibitem{frolov2002}{\it Frolov A., Kofman L. and Starobinsky A.} Prospects and problems of tachyon matter cosmology // Physics Letters B. -- 2002. --  V. 545. -- P. 8--16.
\bibitem{gibbons2003}{\it Gibbons G. W.} Thoghts on tachyon cosmology // Classical and Quantum Gravity. -- 2003. -- V. 20. -- P. S321--S346.
\bibitem{hinshaw2008}{\it Hinshaw G., Weiland J. L., Hill R.S., et al .} Five-Year Wilkinson Microwave Anisotropy Probe (WMAP) Observations: Data Processing, Sky Map \& Basic Results // arXiv: 0803.0732 [astro-ph].
\bibitem{hu1998}{\it Hu W.} Structure Formation with Generalized Dark Matter // Astrophys. J.--1998.--V. 506, No 2.-- P.485-494.
\bibitem{hu1999}{\it Hu W., Eisenstein D.} The Structure of Structure Formation Theories //  Phys. Rev. D.--1999.--V. 59.--id. 083509.
\bibitem{kodama1984}{\it Kodama H., Sasaki M.} Cosmological perturbation theory // Progress of Theor. Phys. Suppl. -- 1984. -- V. 78. -- P. 1--166.
\bibitem{komatsu2008}  {\it Komatsu E., Dunkley J., Nolta M.R., Bennett C.I. et al .} Five-Year Wilkinson Microwave Anisotropy Probe (WMAP) Observations: Cosmological interpretation //  arXiv: 0803.0547 [astro-ph].
\bibitem{kul07}{\it Kulinich Yu., Novosyadlyj B., Pelykh V.} The development of spherically symmetrical perturbation in cosmological models with dark energy. An approximation of the ideal liquid.// Journal of Physical Studies.--2007.--V. 11, No 4.--P.~473--480.
\bibitem{kul08}{\it Kulinich Yu.} Evolution of the spherically symmetric dust-like cloud in $\Lambda$CDM models.// Kinematics and Physics of Celestial Bodies.--2008.--V. 24, No 3.--P.~169--185.
\bibitem{liu2004}{\it Liu J.} Sensitivity of quintessence perturbations to initial conditions // Phys. Rev. D.--2004.--V. 69.-- id. 083504.
\bibitem{lokas02}{\it Lokas E.} Structure Formation in the Quintessential Universe // Acta Physica Polonica B.--2001.--V. 32.--P.3643-3654.
\bibitem{ma1995}{\it Ma C.-P., Bertschinger E.} Cosmological perturbation theory in the synchronous and conformal newtonian gauges // Astrophys. J. -- 1995. -- V. 455. -- P. 7--25. 
\bibitem{Maor05} {\it Maor~I., Lahav~O.} On virialization with dark energy//
Journal of Cosmology and Astroparticle Physics,--2005. --V.2005, Issue 07. --P.003(1--14).
\bibitem{Mota04} {\it Mota~D.F., van de Bruck~C.} On the spherical collapse model in dark energy cosmologies//
Astronomy and Astrophysics, --2004. --V. 421. --P.71-81.
\bibitem{nolta2008}  {\it Nolta M.R., Dunkley J., Hill R.S., et al .} Five-Year Wilkinson Microwave Anisotropy Probe (WMAP) Observations: Angular Power spectra //  arXiv: 0803.0593 [astro-ph].
\bibitem{novosyadlyj2007} {\it Novosyadlyj B.} Formation of the large-scale structure of the Universe: theory and observations // Journal of Physical Studies.--2007.--V. 11, No 2.--P. 226--257.
\bibitem{nunes2006} {\it Nunes~N.J., Mota~D.F.} Structure Formation in Inhomogeneous Dark Energy Models // Mon. Not. Roy. Astron. Soc.--2006.--V. 368.--P. 751-758. 
Astronomy and Astrophysics, --2004. --V. 421. --P.71-81.
\bibitem{padmanabhan2002}{\it Padmanabhan T.} Accelerated expansion of the universe driven by tachyonic matter // Phys. Rev. D.--2002.--V. 66.--id. 021301.
\bibitem{roy-choudhury2002}{\it Padmanabhan T., Roy Choudhury T.} Can the clustered dark matter and the smooth dark energy arise from the same scalar field? // Phys. Rev. D. -- 2002. -- V. 66. -- id. 081301. 
\bibitem{peebles2003} {\it  Peebles P. J. E., Ratra B.} The Cosmological Constant and Dark Energy // Rev. Mod. Phys.--2003.--V. 75.--P. 559--606.
\bibitem{sahni2003} {\it Sahni V., Saini T. D., Starobinsky A. A., Alam U.} Statefinder - A New Geometrical Diagnostic of Dark Energy // J. Exp. Theor. Phys. Lett.--2003.--V. 77.--P. 201--209.
\bibitem{sen2003}{\it Sen A.} Remarks on tachyon driven cosmology // Physica Scripta. -- 2005. -- V. T117. -- P. 70--75.
\bibitem{sergijenko2008}{\it Sergijenko O., Novosyadlyj B.} Scalar field as dark energy accelerating expansion of the Universe // Kinematics and Physics of Celestial Bodies.--2008.--V. 24, No 5.--P. 345--359.
\bibitem{tegmark2006}{\it Tegmark M., et al.} Cosmological constraints from the SDSS luminous red galaxies // Phys. Rev. D. -- 2006. -- V. 74. -- id. 123507.
\bibitem{unnikrishnan2008}{\it Unnikrishnan S., Jassal H.K., Sechadri T.R.} Scalar field dark energy perturbations and their scale dependence //
arXiv: 0801.2017 [astro-ph].
\bibitem{unn2008}{\it Unnikrishnan S.} Can cosmological observations uniquely determine the nature of dark energy? // Phys. Rev. D. -- 2008. -- V.78. -- id.063007.
\bibitem{wang98}{\it Wang L., Steinhardt P.J.} Cluster Abundance Constraints for Cosmological Models with a Time-varying, Spatially Inhomogeneous Energy Component with Negative Pressure // Astrophys. J.--1998.--V. 508, No 2.-- P.483-490.
\bibitem{cmbfast1999} {\it Zaldarriaga~M., Seljak~U.} CMBFAST for spatially closed universes.--1999.--Astrophys. J. Suppl. Series.--V. 29.--No2.--P.431-434. 
\end{thebibliography}
\end{document}